\documentclass[5p,compress]{elsarticle}

\date{February 14, 2022}
\journal{arXiv}
\usepackage{xcolor}
\usepackage{newtxtext}
\usepackage[varvw,bigdelims]{newtxmath}
\usepackage[colorlinks=true,allcolors=blue]{hyperref}
\usepackage{lineno}
\modulolinenumbers[5]

\usepackage{amsmath}

\newcommand{\HHT}[1]{\textcolor{red!70!black}{#1}}
\newcommand{\ONO}[1]{\textcolor{blue!50!gray}{#1}}

\begin{document}

\begin{frontmatter}

\title{Reply to Comment on ``Phase-space consideration on barrier transmission in a time-dependent variational approach with superposed wave packets''}
\author{Akira Ono}
\ead{ono@nucl.phys.tohoku.ac.jp}
\address{Department of Physics, Tohoku University, Sendai 980-8578, Japan}

\begin{abstract}
  In the paper by Hasegawa, Hagino and Tanimura (HHT) [\href{https://doi.org/10.1016/j.physletb.2020.135693}{Phys.\ Lett.\ B 808 (2020) 135693}, \href{https://arxiv.org/abs/2006.06944}{arXiv:2006.06944}], they concluded that quantum tunneling was simulated by a time-dependent generator coordinate method (TDGCM). In contrast, difficulties of TDGCM in describing quantum tunneling were pointed out by Ono in the paper [\href{https://doi.org/10.1016/j.physletb.2022.136931}{Phys.\ Lett.\ B 826 (2022) 136931}, \href{https://arxiv.org/abs/2201.02966}{arXiv:2201.02966}]. Recently, HHT submitted a comment [\href{https://arxiv.org/abs/2202.00513v1}{arXiv:2202.00513v1}], by which they appear to give some counterarguments to Ono's paper. Here I examine their arguments, reviewing the main points of Ono's paper. The confusion in HHT's arguments may be mostly resolved by recognizing how the wave packets interfere with each other when they are coherently superposed.
\end{abstract}

\end{frontmatter}

\section{Background}

In Ref.~\cite{hasegawa2020}, Hasegawa, Hagino and Tanimura (HHT) employed a time-dependent generator coordinate method (TDGCM) aiming at a description of quantum tunneling. Although the final goal is to establish a method for many-particle problems, the illustrative example studied in their article \cite{hasegawa2020} is equivalent to a problem of one particle in one dimension. Their conclusion was that quantum tunneling could be simulated by TDGCM.

On the other hand, in Ref.~\cite{ono2022}, Ono investigated the same problem under the same TDGCM model, by paying attention to the time evolution of the phase space distribution, which is easy for such an illustrative problem of one particle in one dimension. In particular, it was noticed that the free propagation of the incoming state is already non-trivial in the TDGCM solution, which affects how the potential barrier is passed over. One of the main conclusions was that the barrier passage can be of classical nature due to the high-momentum components in the prepared initial state. Such classical barrier passage cannot be regarded as quantum tunneling, contrary to the conclusion of Ref.~\cite{hasegawa2020}. Based on the understanding in the phase space, the reason why HHT reached the different conclusion was explained in Ref.~\cite{ono2022} in the form of comments or criticisms on Ref.~\cite{hasegawa2020}.

Now a comment to Ref.~\cite{ono2022} has been submitted to arXiv by HHT \cite{hasegawa2022v1}, in which they apparently give counterarguments to Ono's comments in Ref.~\cite{ono2022}. My comments on their individual arguments are given below in this article. However, I first point out here that their arguments \cite{hasegawa2022v1} do not take into account the main point of Ref.~\cite{ono2022}, which is the consideration based on the phase space distribution. Without that consideration, it is hardly possible to understand the behavior of the solution of TDGCM. Of course, I will try to follow HHT's arguments as far as possible, but it is sometimes necessary to review the phase-space consideration in Ref.~\cite{ono2022}.

In the first paragraph of Ref.~\cite{hasegawa2022v1}, Ono's comments in Ref.~\cite{ono2022} are mentioned by HHT as ``\HHT{The claim is based on the fact that the initial wave function employed in Ref.~\cite{hasegawa2020} has a broad momentum distribution}.'' This statement is not rigorously correct because Ono's claim is based on the observed time evolution of the phase space distribution, i.e., the solution of the TDGCM equation. What is directly affected by the initial broad momentum distribution is the solution of TDGCM, without any room for personal opinions to enter.

\section{Does the momentum width play any role?}

\begin{figure*}
\begin{minipage}[b]{0.5\textwidth}\centering
\includegraphics[width=0.7\textwidth]{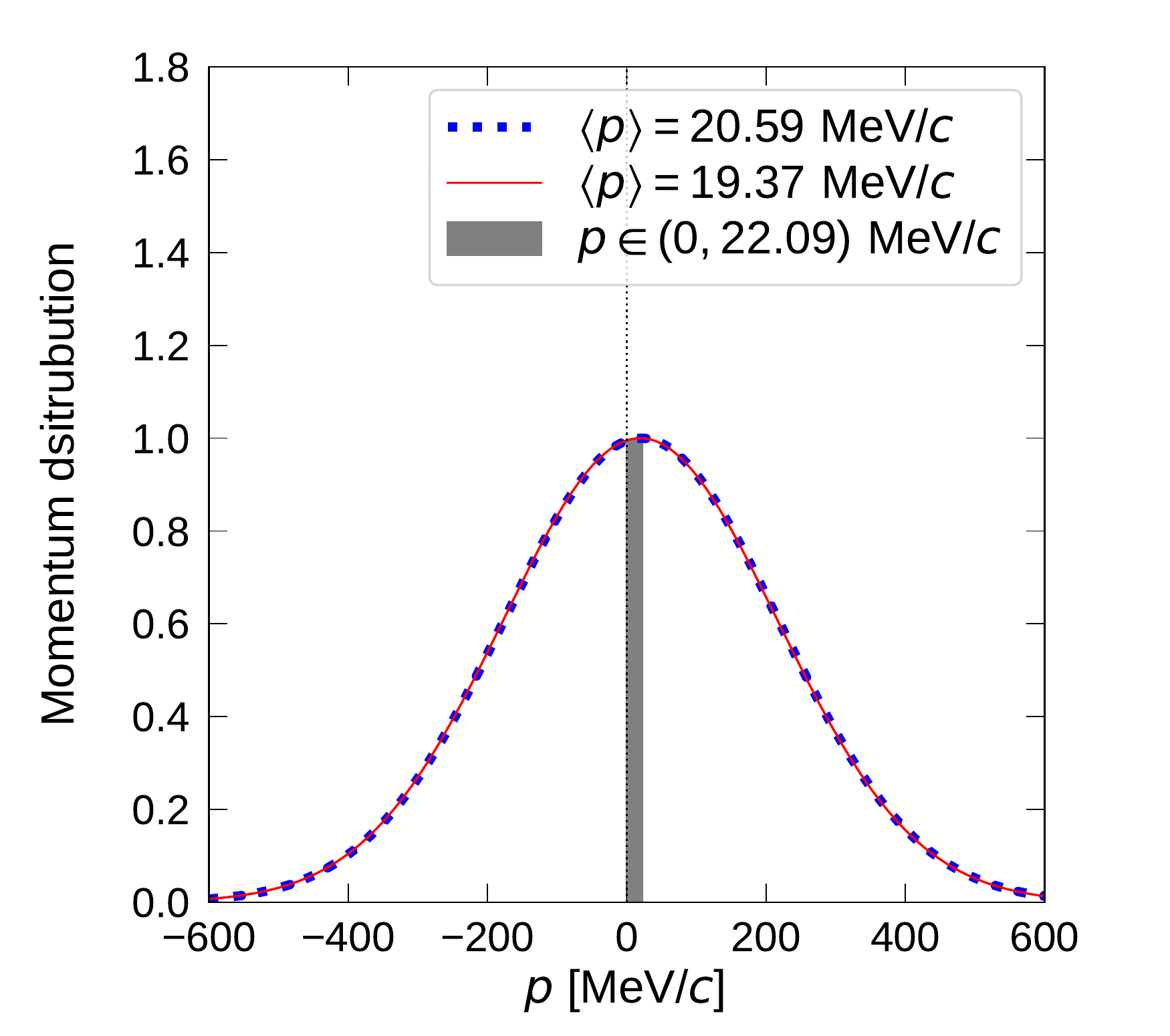}
\end{minipage}
\begin{minipage}[b]{0.5\textwidth}
\caption{\label{fig:momdst}
The momentum distributions for the two Gaussian wave packets which were used in the initial state in the calculation of Ref.~\cite{hasegawa2020}. The gray area indicates the region that is relevant to quantum tunneling ($E<0.13$ MeV and $p>0$). This figure is Fig.~1 of Ref.~\cite{ono2022}.
}
\bigskip
\end{minipage}

\bigskip\bigskip
\includegraphics[width=\textwidth]{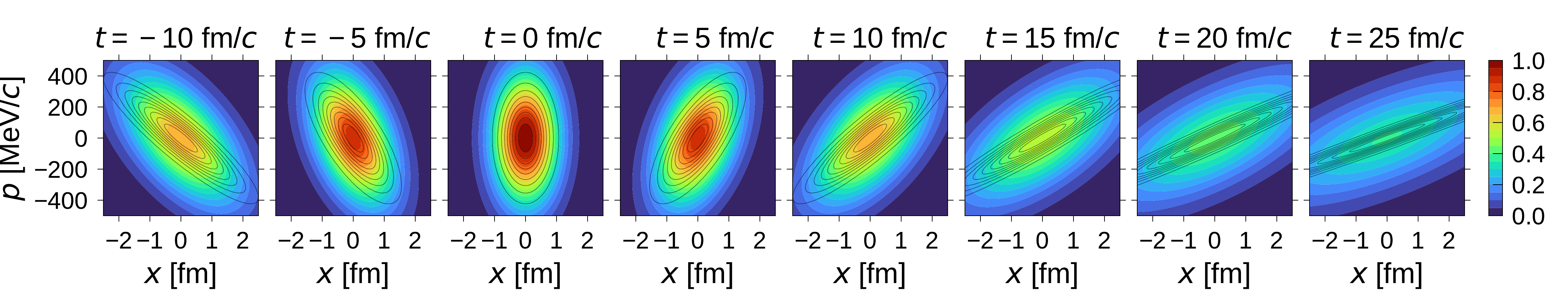}
\caption{\label{fig:freef}
Wigner and Husimi functions for a free particle motion, at different times from $t=-10$ fm/$c$ to 25 fm/$c$. The Wigner function is shown by the contour lines drawn for $f=0.2, 0.4, 0.6,\ldots$, while the Husimi function is represented by the color scale. The initial condition for the Wigner function was chosen as $f_{\text{W}}(x,p,t=0)=2\exp(-2\nu_rx^2-p^2/2\hbar^2\nu_r)$ which is very similar to the Wigner function of the initial state $\psi(x,t=0)$ of the TDGCM calculation in Ref.~\cite{hasegawa2020}. This figure is Fig.~2 of Ref.~\cite{ono2022}.
}

\bigskip\bigskip
\begin{minipage}[b]{0.5\textwidth}\centering
\includegraphics[width=0.9\textwidth]{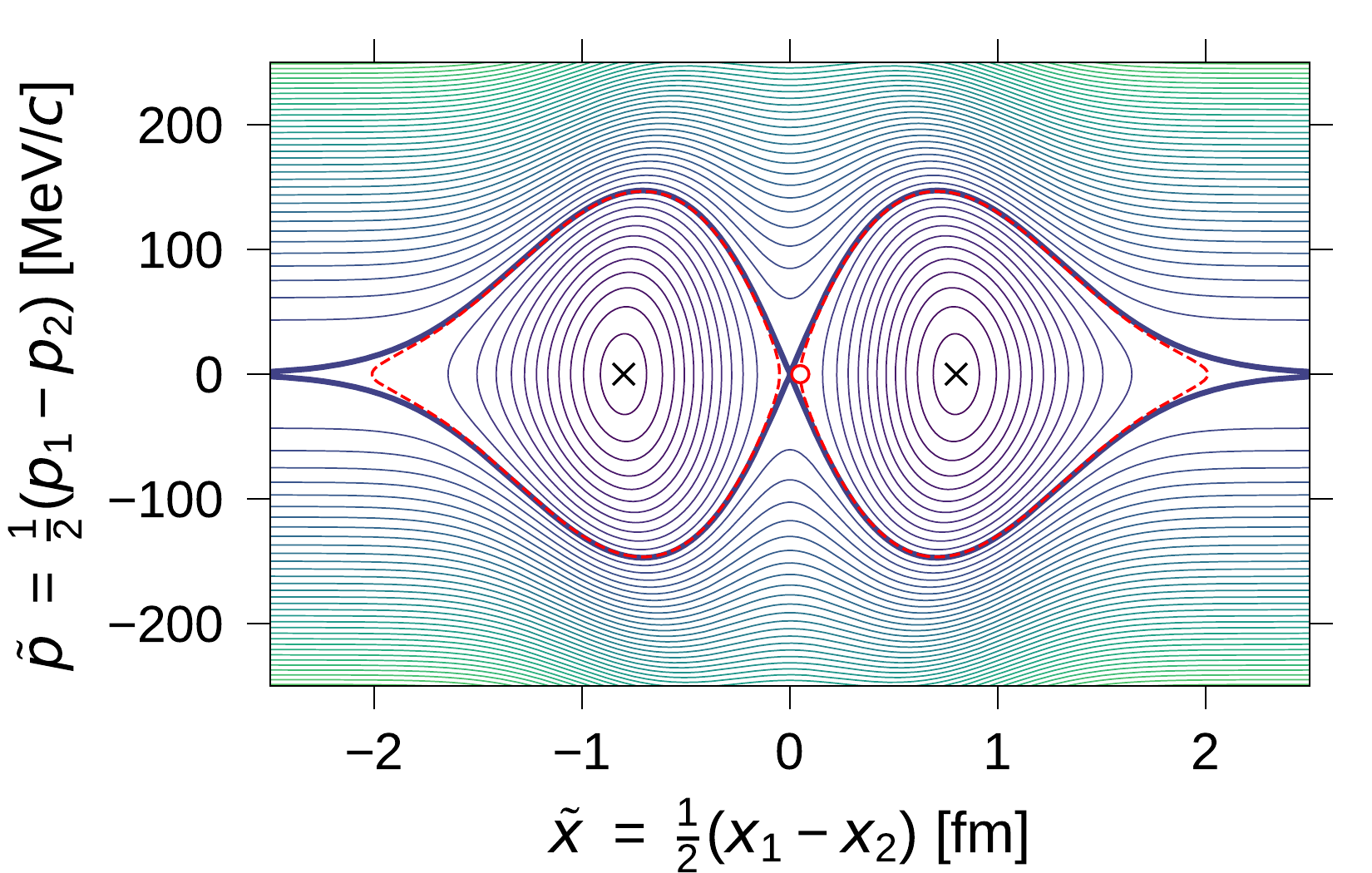}
\end{minipage}
\begin{minipage}[b]{0.5\textwidth}
\caption{\label{fig:ekincontour} Contour plot for the energy $E$ of a free particle when the wave function is approximated by a superposition of two Gaussian wave packets, centered at $(\tilde{x},\tilde{p})$ and $(-\tilde{x},-\tilde{p})$, with equal coefficients that are in phase.  Contour lines are drawn for $E=\hbar^2\nu_r/2\mu + k\times 0.5\ \text{MeV}$ for integers $k$. The thick contour line is for $E=\hbar^2\nu_r/2\mu=10.38$ MeV.  The energy takes the minimum $E=4.60$ MeV at $(\tilde{x}, \tilde{p})=(\pm0.80\ \text{fm}, 0)$, shown by the crosses. The open circle indicates the initial condition chosen in Ref.~\cite{hasegawa2020}, $(\tilde{x}, \tilde{p})=(0.05\ \text{fm},\ 0.61\ \text{MeV}/c)$. The dashed contour line is for $E= 10.33$ MeV corresponding to this initial condition. The width parameter of the Gaussian wave packets and the mass are chosen to be $\nu_r=1\ \text{fm}^{-2}$ and $\mu=1876$ MeV. This figure is Fig.~3 of Ref.~\cite{ono2022}.}
\bigskip
\end{minipage}

\bigskip\bigskip
\includegraphics[width=\textwidth]{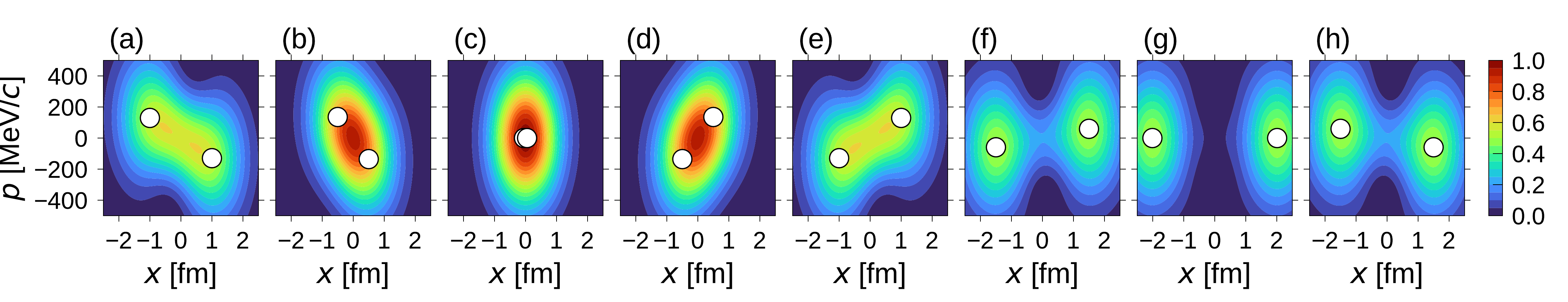}
\caption{\label{fig:husimif}
Husimi function for the TDGCM state of Eq.~(\ref{eq:gcmwf}) in a boosted frame. Each panel represents the Husimi function by the color scale, for the state specified by the parameters $(\tilde{x}, \tilde{p})$, whose values are chosen on the dashed contour line in Fig.~\ref{fig:ekincontour}. The open circles represent the locations of the wave packet centers, $\pm(\tilde{x},\tilde{p})$.
This figure is Fig.~4 of Ref.~\cite{ono2022}.
%
%
}
\end{figure*}

In the second paragraph of Ref.~\cite{hasegawa2022v1}, HHT agree that the illustrative problem treated in Refs.~\cite{hasegawa2020, ono2022} is equivalent to one particle in one dimension, in which the TDGCM model expresses the wave function (for the relative motion of the two colliding nuclei) by a superposition of Gaussian wave packets,
\begin{equation}
\psi(x,t)=\sum_af_a(t)\, e^{-\nu_r\bigl(x-\frac{z_a(t)}{\sqrt{\nu_r}}\bigr)^2}.
\label{eq:relativewf}
\end{equation}
Here the wave packet center $z_a(t)$ of each component $a$ takes a complex value, and its real and imaginary parts have information on the spatial and momentum centers, $x_a(t)$ and $p_a(t)$, respectively. The width parameter $\nu_r$ is a fixed constant, e.g., $\nu_r=1\ \text{fm}^{-2}$, which determines the spatial and momentum widths, $\Delta x=1/(2\sqrt{\nu_r})$ and $\Delta p=\hbar\sqrt{\nu_r}$; a Gaussian wave packet has the minimum uncertainty $\Delta x\Delta p=\frac12\hbar$. Figure~\ref{fig:momdst} shows the momentum distributions of the two Gaussian wave packets used in the initial state of the calculation of Ref.~\cite{hasegawa2020}. This is the same figure as Fig.~1 of Ref.~\cite{ono2022}, which is put in this article for the convenience. (The same for the other three figures.)

It was properly mentioned in Ref.~\cite{ono2022} that the width parameter $\nu_r$ is a fixed constant, and therefore the Gaussian wave packets do not change their shape, though they can, of course, move by the changes of $x_a(t)$ and $p_a(t)$. The second paragraph of Ref.~\cite{hasegawa2022v1} seems to be mainly focusing on such time evolution of the individual Gaussian wave packets. However, such a consideration is insufficient to understand the wave function $\psi(x,t)$ of Eq.~\eqref{eq:relativewf}. The effect of interference between Gaussian wave packets needs to be considered, in addition to the ``classical'' motions of the Gaussian wave packets. For example, through the interference, the momentum width can affect the motions of the Gaussian wave packets, as we can find examples in Ref.~\cite{ono2022} and also as reviewed below.

We should keep in mind that the Gaussian wave packets are used as basis functions to express the wave function $\psi(x,t)$ which represents the physical state of the system at the specified time $t$. In Ref.~\cite{ono2022}, the term `wave function' is used for the physical state $\psi(x,t)$, but not for the Gaussian wave packet in each term in the right-hand side of Eq.~\eqref{eq:relativewf}. Generally speaking, one cannot necessarily find physical meaning in individual basis functions. In particular, in TDGCM, the Gaussian wave packets are not orthogonal with each other, which makes it further difficult to find physical meaning in individual superposed components. This was already warned, e.g., in Ref.~\cite{ono2022}.

The wave function $\psi(x,t)$ of Eq.~\eqref{eq:relativewf} for the physical state does not have the same spatial and momentum widths as the individual Gaussian wave packets, not only due to the distribution of the Gaussian centers but also as a consequence of constructive and destructive interference between Gaussian wave packets. As mentioned in Ref.~\cite{ono2022}, Gaussian wave packets can form an overcomplete basis. Therefore, when many Gaussian wave packets are suitably superposed, one can, in principle, reproduce the exact solution of the Schr\"{o}dinger equation which is displayed in Fig.~\ref{fig:freef} for the case of free propagation, in the form of the phase space distribution. Such a time evolution of $\psi(x,t)$ is consistent with the idea of HHT \cite{hasegawa2022v1} on ``\HHT{a quantum mechanical wave packet, whose spatial width broadens as a function of time according to its momentum distribution}.'' Note that, in this ideal case, the momentum distribution in the initial state is fully reflected in the time evolution of $\psi(x,t)$ in TDGCM. This is therefore a counter-example to HHT's assertion \cite{hasegawa2022v1} ``\HHT{even if the wave function \eqref{eq:relativewf} may have a broad momentum distribution, that will not be reflected fully in the time-dependent dynamics}.''

\section{Why are trajectories entangled?}

Because of the interference between Gaussian wave packets, the time evolution of the wave packet centers, $x_a(t)$ and $p_a(t)$, is highly non-trivial, as demonstrated for the free propagation in Ref.~\cite{ono2022}, in the case of two wave packets superposed with equal weights. Figure \ref{fig:ekincontour} shows the energy contour lines in the boosted frame in which $(x_1(t),p_1(t))=-(x_2(t),p_2(t))\equiv(\tilde{x}(t),\tilde{p}(t))$. A solution of the TDGCM equation follows one of the contour lines, e.g., cyclically on the red dashed line in the case of the calculation of Ref.~\cite{hasegawa2020}. The momentum centers $p_a(t)$ are changing quite a lot, of the order of 100 MeV/$c$, already in this free propagation without any effect of the potential, which is, of course, counter-intuitive since one would naturally expect that the momentum and the energy should be conserved for the free propagation. I do not know whether HHT have accepted this fact of the drastic change of $p_a(t)$ during the free propagation.

Such non-trivial motions of the wave packet centers may be considered as due to some entanglement between different trajectories, e.g., $(x_1(t),p_1(t))$ and $(x_2(t),p_2(t))$, as mentioned by HHT in the third paragraph of Ref.~\cite{hasegawa2022v1}. They think, ``\HHT{the idea of the time-dependent generator coordinate method \eqref{eq:relativewf} is nothing more than that of the entangled trajectory molecular dynamics (ETMD) developed in quantum chemistry \cite{donoso2001,donoso2003}}.'' In my opinion, this analogy to ETMD needs to be taken with cautions because there are some differences between TDGCM and ETMD. In ETMD, the test particles are used as elements to represent the phase space distribution $f(x,p,t)$. Each test particle is localized in a small region in the phase space, and $f(x,p,t)$ is expressed as a simple summation of the positive contributions from test particles. On the other hand, in TDGCM, each wave packet occupies a relatively broad region in the phase space determined by the minimum uncertainty $\Delta x\Delta p=\frac12\hbar$, and the contribution to $f(x,p,t)$ is a complicated summation of complex values due to interference between wave packets. For example, see Eq.~(12) of Ref.~\cite{ono2022} for TDGCM, which is in stark contrast to Eq.~(5) of Ref.~\cite{donoso2001} for ETMD. Furthermore, the entanglements are not from the same origin, e.g., as we can see that free propagation is described in ETMD naturally with no entanglement occurring, while the wave packets are non-trivially entangled in TDGCM already for free propagation.

Nevertheless, HHT argues \cite{hasegawa2022v1}, ``\HHT{In ETMD, a quantum tunneling is simulated with many classical trajectories by making them entangled during the time evolution. This corresponds to taking a linear superposition of many Slater determinants in the time-dependent generator coordinate method. Such entanglement is a key to recover the quantum mechanical behavior of time-dependent dynamics}.'' However, if we consider the cautions mentioned above, the partial success of ETMD for quantum tunneling does not guarantee an automatic success of TDGCM for the same problem. As already commented in Ref.~\cite{ono2022}, the results of TDGCM have to be carefully analyzed, e.g., similarly to what has been done for ETMD or as suggested by Ref.~\cite{ono2022}. Nothing has been reported on such tests by HHT \cite{hasegawa2022v1}.

It is still instructive to see what is causing the counter-intuitive entanglement in TDGCM for free propagation, which was already argued in Ref.~\cite{ono2022}. The phase-space distribution function (Husimi function) was displayed in Fig.~\ref{fig:husimif}, for the TDGCM states corresponding to eight points selected on the red dashed contour line in Fig.~\ref{fig:ekincontour}. We can see in Fig.~\ref{fig:husimif} that the evolution from the panel (b) to (d) is quite similar to the exact time evolution displayed in Fig.~\ref{fig:freef} from $t=-5$ to 5 fm/$c$, i.e., the TDGCM with two wave packets provides an accurate description for this short time interval. The time evolution of the phase space distribution can be easily understood based on the free propagation of each phase space element. Being induced by the momentum distribution, a correlation develops between the position and the momentum in the phase space. In Fig.~\ref{fig:husimif}, the Gaussian centers are indicated by open circles. From the state in panel (c) to (d), the momentum centers change a lot from $p_a=0$ to $p_a\sim\pm 150$ MeV/$c$. Such a change of momenta during a short time of 5 fm/$c$ may be counter-intuitive as a free propagation, but it provides the natural time evolution of the phase space distribution and therefore the wave function $\psi(x,t)$. Thus, the trajectories of Gaussian centers may be interpreted as entangled by struggling to reproduce the free propagation, not of the Gaussian centers themselves but of the phase space distribution and $\psi(x,t)$, after complicated interference between them. This is also an example to show that the momentum distribution in $\psi(x,t)$ plays an important role in the time evolution in TDGCM.

It is noteworthy that the phase space distribution is not a simple summation of two Gaussian functions in the state of panel (d), for example. While the total distribution is spreading in a tilted direction in the phase space, it is shrinking in the other direction, compared to the original width at the state of (c). Such a shrinking can be achieved only through interference. The same is true, e.g., for the minimum energy configuration in Fig.~\ref{fig:ekincontour} shown by the cross symbols, at which the momentum width is considerably narrower than the original width $\Delta p=\hbar\sqrt{\nu_r}$ of the Gaussian wave packet \cite{ono2022}.

Thus, the effort of superposing wave packets is spent already for the free propagation of the incoming state when TDGCM is employed for a collision problem as in Ref.~\cite{hasegawa2020}. This is in contrast to ETMD in which entanglements occur to describe purely quantum effects. Furthermore, in TDGCM, the description of free propagation is good only during a limited time interval, and the time evolution is pathological at most of the times as pointed out by Ref.~\cite{ono2022}, when only a few wave packets are superposed.

\section{Is TDGCM simulating a fixed beam energy?}

For the TDGCM solution calculated by HHT in Ref.~\cite{hasegawa2020}, it seems they assume that the simulation corresponded to an almost fixed beam energy because $E_1(t_0)=p_1^2(t_0)/2\mu$ and $E_2(t_0)=p_2^2(t_0)/2\mu$ were chosen to be almost identical at the initial time $t_0=0$. Here $\mu$ is the (reduced) mass and that energy was denoted by $E(t=0)\approx E_1(0)\approx E_2(0)$ in Ref.~\cite{ono2022}. This picture based only on the momentum centers $p_a(t)$ is called View B \cite{ono2022}. Roughly speaking, View B ignores the momentum width inherent in the individual wave packets in estimating the energy. Since the energy $E(t=0)=0.11$ MeV was chosen below the barrier height $V_{\text{B}}=0.13$ MeV, any transmission over the barrier was regarded as quantum tunneling in Ref.~\cite{hasegawa2020}.

This interpretation of the simulation was criticized by Ono in Ref.~\cite{ono2022}, as quoted here: ``\ONO{Ref.~\cite{hasegawa2020} took View B, abandoning interpretation of the wave function $\psi(x,t)$. They chose the condition at $t=0$ so that all $p_a^2(t=0)/2\mu$ have almost the same value $E(t=0)$, and the simulation was regarded as corresponding to this beam energy. They then claimed that barrier transmission was all due to quantum tunneling because $E(t=0)<V_{\text{B}}$. This is however not correct. What one would claim depends on what time was regarded as the `initial' time $t_0$, because $p_a(t_0)$ depend on $t_0$ without any effect of the potential in the incoming state. It is actually impossible to simulate a fixed beam energy in TDGCM. There is no foundation for View B with $E(t=0)$ regarded as the beam energy, when wave packets are coherently superposed. We can also expect that the result of a simulation depends on the timing of arrival at the barrier, which makes any statement questionable if it is based on only a single simulation}.'' 

The fourth paragraph of Ref.~\cite{hasegawa2022v1} by HHT might be a counterargument to this criticism. However, none of the points in Ono's criticism is rebutted by HHT. If they were to persist in View B to regard $E(t=0)$ as the beam energy and to justify their result \cite{hasegawa2020} as an indication of quantum tunneling, they needed to have given counterarguments to individual points in Ono's criticism quoted above. This was not done in Ref.~\cite{hasegawa2022v1}.

Instead, HTT \cite{hasegawa2022v1} assert their opinion as ``\HHT{In our opinion, the energy spreading inherent in each Gaussian wave function should not fully be included in estimating the total energy spreading, as it does not contribute to the dynamics when a single wave packet is involved. That is, one should make a clear distinction between the energy spreading inherent in single-particle wave functions and that due to a distribution of initial conditions. As the former does not contribute to the dynamics when a single wave packet is considered, it is not trivial to estimate appropriately the energy spreading of initial wave packet for the TDGCM when a Gaussian function with a fixed width is employed}.'' Although this statement by HHT is a little vague because of the usage of ``\HHT{not fully}'' and ``\HHT{not trivial},'' their opinion seems to adopt View B to regard the distribution of $E_a(t_0)=p_a^2(t_0)/2\mu$ as that of the beam energy. As mentioned by Ono \cite{ono2022}, when the model uses only a single wave packet, we know many cases in which View B works even though it is not fully consistent with quantum mechanics. HHT misunderstand that the same is true for TDGCM, by which they draw an incorrect conclusion. As already demonstrated in previous sections, we have to consider the effect of the interference between the superposed wave packets, which makes it meaningless to divide the distribution into a frozen part and the rest. Also, as quoted above from Ref.~\cite{ono2022}, their opinion indeed includes serious inconsistencies as a practical problem.

In some sense, it is true that the momentum width is reflected only partially in the dynamics when only two Gaussian wave packets are superposed in the setup of Ref.~\cite{hasegawa2020}. This is because of the pathological evolution observed in Fig.~\ref{fig:husimif}, and therefore shows a defect of the model rather than its ability to predict barrier transmission in a controlled way. The argument in Ref.~\cite{ono2022} is quoted here: ``\ONO{The behavior of the incoming state is already nontrivial as discussed in the previous section, and the cyclic motion in the incoming state is very slow in the pathological phase [from (e) through (g) to (a) in Fig.~\ref{fig:husimif}]. Therefore, the potential barrier will be encountered with a large chance in the pathological phase. For example, if the barrier is encountered in the stage of (f), the front part of the distribution will pass over the barrier when the average momentum $\langle p\rangle$ in this part is high enough. In fact, each of the front and rear parts is similar to a Gaussian wave packet like the initial state in panel (c), and even a slightly larger $\langle p\rangle$ than the initial one may be sufficient for barrier passage (as reminded by Fig.~\ref{fig:momdst}). Such a barrier passage by a part with a frozen shape is not called tunneling in general. The same can occur in usual AMD and TDHF when $\langle p\rangle$ is high enough. This consideration on the above example seems to be consistent with the case shown in Fig.~2 of Ref.~\cite{hasegawa2020}}.''

\section{Can energy conservation be violated?}

Quantum tunneling is a phenomenon that cannot occur in the corresponding classical system under the condition of energy conservation. It is, therefore, hardly possible to argue quantum tunneling if the model does not have well-defined conserved energy. As pointed out by Ono \cite{ono2022}, when barrier transmission occurred, the observed energy in the transmission (and also reflection) channel was significantly different from the beam energy $E(t=0)$ in the TDGCM solution of Ref.~\cite{hasegawa2020}. Although it may depend on the source of the problem, we should not regard any model with such energy non-conservation as qualified for studying quantum tunneling.

Now HHT argue in the fifth paragraph of Ref.~\cite{hasegawa2022v1}. For the interpretation about the seriousness of the energy non-conservation mentioned above, they stated, ``\HHT{a care must be taken in this interpretation, due to the well known spurious cross-channel correlation in the time-dependent Hartree-Fock theory. This problem may not be cured completely if the number of Slater determinants superposed in the wave function is small}.'' I agree with this understanding for TDHF; this is indeed why TDHF cannot describe quantum tunneling. The problem can be considered as the unphysical nonlinearity caused by the limitation of the wave function. The problem in TDHF is usually argued in many-body systems, while TDGCM with a limited number of Gaussian wave packets suffers from this problem already in one-particle systems.  HHT seem to agree now that TDGCM does not conserve energy due to a well-known fundamental defect of this kind of model, and therefore it is not qualified yet as a model for quantum tunneling.

It is difficult to grasp the final sentence of HHT, ``\HHT{As in ETMD, only the total energy has a clear physical meaning in considering quantum tunneling, rather than the energy of each `test particle'}.'' In the present context, the only possible interpretation of the ``\HHT{total energy}'' is the sum of the kinetic and potential energies, since the Hamiltonian of this one-particle system has only these two terms. The same word is used for this meaning, e.g., in some papers of ETMD \cite{donoso2003,aswang2009,lfwang2012}, where each test particle has its total energy, in contrast to HHT's sentence. The total energy of each test particle does not need to be conserved, which, however, does not have a direct physical meaning. The test particles represent the phase-space distribution function that presumably corresponds to a quantum state. Such a quantum state treated in phase-space approaches is usually not an energy eigenstate, which we should interpret as such. In the asymptotic region where the potential is a constant, it should be straightforward, in principle, to extract the distribution of the energy from test particles in each channel. 

In case the ``\HHT{total energy}'' means the energy averaged over all channels, i.e., the energy expectation value for $\psi(x,t)$, the statement by HHT implies that ETMD is also not qualified, but we need to check the literature. In Refs.~\cite{aswang2009, lfwang2012}, repaying the borrowed energy by a transmitted test particle is argued, with which one may expect that the relation of the energies of different channels could be suitably described by ETMD. Some practical tests have been performed for ETMD in the literature, but it does not guarantee an automatic success of TDGCM. As mentioned before and in Ref.~\cite{ono2022}, similar and other tests are necessary for TDGCM to claim quantum tunneling.

\section{Summary}

Individual comments from HHT \cite{hasegawa2022v1} were examined. Since HHT did not consider the main point of Ono's paper \cite{ono2022}, necessary reviews were also made here. Probably, the confusion in HHT's arguments can be mostly resolved if they understand how the Gaussian wave packets interfere with each other. After all, there is nothing to change in Ono's paper \cite{ono2022}, including the criticisms to the conclusion of HHT's paper \cite{hasegawa2020}.

\section*{Acknowledgments}
This work was supported by JSPS KAKENHI Grant Numbers JP17K05432 and JP21K03528.

\bibliography{ono_nucl}

\end{document}